\documentclass[twocolumn,prl,showpacs]{revtex4}

\usepackage{graphicx}
\usepackage{rotating}
\usepackage{amsmath}
\usepackage{amsfonts}
\usepackage{amssymb}
\usepackage{enumerate}
\usepackage{longtable}
\usepackage{dcolumn}
\usepackage{bm}

\newcommand{\be}{\begin{equation}}
\newcommand{\ee}{\end{equation}}
\newcommand{\bn}{\begin{eqnarray}}
\newcommand{\en}{\end{eqnarray}}

\def\thb2{\frac{\theta_{i}}{2}}

\def\gdi{$GdI_2$}

\def\dz2{$d_{z^2}$}
\def\dx2y2{$d_{x^2-y^2}$}
\def\dxy{$d_{xy}$}

\begin{document}

\author {A. Taraphder$^{1,4}$, M. S. Laad$^{1}$, L. Craco$^{2}$,
and A. N. Yaresko$^{3}$ }
\title {\gdi: A New Ferromagnetic Excitonic Solid?  }
\affiliation{ $^{1}$Max-Planck-Institut f\"ur Physik Komplexer Systeme,
01187 Dresden, Germany \\
$^{2}$Max-Planck-Institut f\"ur Chemische Physik
Fester Stoffe, 01187 Dresden, Germany \\
$^{3}$Max-Planck-Institut f\"ur Festkoerperforschung,
70569 Stuttgart, Germany \\
$^{4}$Department of Physics and  Centre for
Theoretical Studies,\\ Indian Institute of Technology, Kharagpur 721302 India }

\date{\rm\today}

\begin{abstract}
The two-dimensional, colossal magnetoresistive system \gdi\, develops
an unusual metallic state below its ferromagnetic transition and
becomes insulating at low temperatures. We argue that this geometrically
frustrated, correlated poor metal is a possible candidate for
a ferromagnetic excitonic liquid. The renormalized Fermi surface supports
a further breaking of symmetry to a charge ordered, excitonic
solid ground state at lower temperatures via order by disorder mechanism.
Several experimental predictions are made to investigate this unique orbitally
correlated ground state.

\end{abstract}

\maketitle

Correlated electronic systems with geometric frustration have spawned 
intense interest recently, because of the startling diversity of physical 
properties they manifest. The crucial roles of electronic correlation and 
frustration, working in tandem, are nevertheless poorly understood from a 
fundamental viewpoint: strong correlations generically drive Mott 
transitions (accompanied by orbital/magnetic order) and manifest spectacular 
changes in physical responses under minute perturbations \cite{IMA_RMP}. 
In correlated systems, geometric frustration gives rise to 
large, exponential degeneracy
of {\it classical} ordered states, and to unconventional order via
an {\it order by disorder} mechanism \cite{DIEP}. The variety of new,
hitherto unexplored, phases of matter emerging in systems where {\it both} 
correlation and frustration are operative is a fascinating, more-or-less 
open problem.

The layered (in XY plane) ferromagnetic (FM) system \gdi, studied by several
authors \cite{FAK+99,AFK+00,ETF+01,TM4}, may turn out to be an
interesting example. Gd ions in the $4f^{7}5d^{1}$ state
form a triangular lattice, separated by iodine layers.
Itinerant electronic states in the partially filled
$d$-bands coupled to localized $f$-electrons
drive an FM state ($T_{c}\simeq 300$~K).
Interestingly, \gdi\, shows significant magnetoresistance \cite{FAK+99} 
(40\% per Tesla) close to $T_{c}$\, \cite{KAS84}. 
In addition, it is iso-structural to the well known dichalcogenides with
hexagonal layered structure \cite{AEBI01}, showing  
charge density wave (CDW) order at low temperatures. In fact,
\gdi\, is a bad metal below $T_{c}$ and becomes insulating
for $T<80$~K. Might this be an indication
that the features seen in dichalcogenides also occur in \gdi?

The unusually high (significantly higher than $\rho_{Mott}^{max}$) resistivity 
of \gdi\, even in the metallic state is strongly at odds with a simple 
FM $s(d)-f$ lattice model, where FM order implies a good metal 
for a translationally invariant system. In fact, experiments yield 
$k_{F}l<1$ ($l$ is electron mean-free path) in \gdi, indicating a strong
inelastic scattering of carriers and resulting in an {\it incoherent}
metallic state, without Fermi liquid quasi-particles.
On the other hand, the FM spin fluctuations \cite{DEI04} 
are well-described by a {\it classical} picture of two-dimensional (2D) FM 
Heisenberg model. The saturation magnetic moment in the FM state is about 
7.33$\mu_B$ per $Gd$ atom, considerably lower than the maximum attainable 
value ($8\mu_B$), indicating the relevance of the competition between strong 
correlation effects and itinerancy among the $5d$ electrons of $Gd$. The 
above observations force one to seek additional, non-magnetic, strong 
correlation effects {\it deep} in the FM state to reconcile magnetism 
with transport in \gdi. Moreover, owing to its $2D$ structure, the role of 
correlation is expected to be very important in \gdi. Strong magnetic
fluctuations found experimentally \cite{RYA05} and a series of magnetic 
ground states obtained in a correlated electronic model \cite{TM4}, 
also underline the relevance of $Gd\,d$-shell correlation.  

Band structure results for \gdi\, \cite{FAK+99,TM4} indicate a spin 
splitting of the conduction band. The crystal field in \gdi\, is relatively 
small and the 5$d$\, orbitals are weakly split and partially occupied.
Emergence of an insulating ground state in with three partially
filled $d$-bands is surprising, since the band-filling is off-commensurate
(less than $1/3$ filling).  This suggests that the $T\rightarrow 0$
insulator must be associated with an additional broken symmetry.  The
high-$T$ bad-metal phase implies that this must emerge
from an instability of an incoherent metallic state, pointing to  
the strong correlation limit. This raises deeper questions:
(i) what is the specific nature of the inelastic scattering processes 
leading to the ``bad metal''?
(ii) How does such an incoherent FM state evolve into a $T\rightarrow 0$
insulator?  Is any additional symmetry breaking (in charge, orbital sectors)
involved?

These observations should go hand-in-hand with the evolution of the
{\it correlated} electronic structure of \gdi\, with temperature. In this 
letter, we address these issues in detail. Using dynamical mean-field
theory (DMFT), we solve a multi-orbital Hubbard model with the LSDA 
band-structure of \gdi\,  \cite{TM4}. We show that strong correlations 
localize two degenerate bands with \dx2y2, \dxy \, orbital character, 
leaving a renormalized band with \dz2 character to cross $E_{F}$.
The renormalized Fermi surface is strongly modified by correlations,
eroding the LSDA hole pockets (see below). Further, we argue that the shape 
of the {\it correlated} Fermi surface {\it naturally}
drives a charge-ordered (CO) ground state with $\sqrt{3}\times
\sqrt{3}$ order on the triangular lattice (breaking the underlying
translational symmetry). Using LSDA+DMFT, we propose a simple
effective model that allows us to suggest experimental signatures of such
a ground state.

The LSDA band structures obtained \cite{TM4} using LMTO method \cite{MAT73} 
shows that the most relevant bands crossing  $E_{F}$ are mainly derived from 
the 5$d$ orbitals (occupations 
of \dx2y2,\,\dxy, \,\dz2 are 0.31, 0.31, 0.30 respectively), split about 
0.9 $eV$ into majority and minority spin sub-bands by strong exchange 
interaction ($J_{fd}$) with the completely spin-polarized 4$f$ shell. The 
corresponding hexagonal Fermi surface, exhibits six hole pockets 
(not shown). 

\begin{figure}[tbp!]
\begin{center}
\includegraphics[width=\columnwidth]{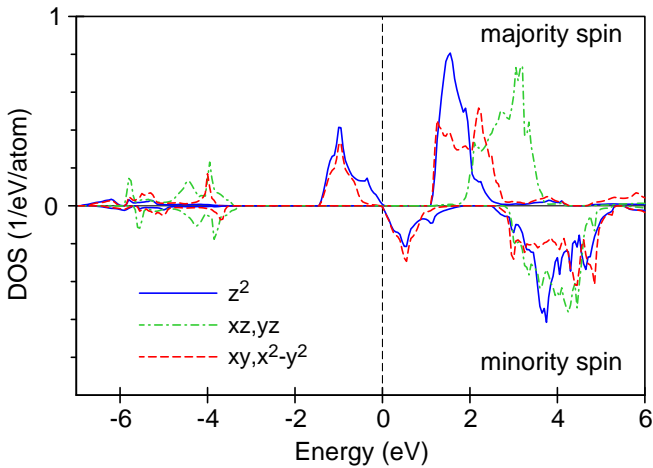}
    \caption{(Color online)
The LSDA+U DOS of the five $Gd \, d-$bands in the majority and minority 
spin sector for $U=6$ eV} 
\label{SP_ud6}
\end{center}
\end{figure}

LSDA+U calculations were performed treating Gd $f$ states as completely 
spin-polarized quasi-core states (U applied only to Gd $d$ states). 
For the double counting term the so-called 
``atomic limit" was used~\cite{LAZ95,YAF03}. We find (Fig.~\ref{SP_ud6}) 
that \dx2y2, \dxy\, and \dz2\, bands are pulled down by {\it static}, orbital 
correlations, and the \dxy, \dx2y2 bands are localized. 
However, LSDA+U {\it overestimates} band-splittings, while 
LSDA cannot split partially occupied bands. A reliable picture requires a 
full incorporation of {\it dynamical} $d$-band correlations, using LSDA+DMFT, 
to which we turn below.

The correlated multi-orbital model Hamiltonian for \gdi\, is described by
$H=H_{0}+H_{int}$, where

\be H_{0}=\sum_{k,a}\epsilon_{a}({\bf k})c_{ka\sigma}^{\dag}c_{ka\sigma}+
\sum_{i,a,\sigma}\Delta_{a}\hat{n}_{ia\sigma}
\ee  
\noindent is the band part with $a,b=$\dz2,\dx2y2,\dxy, and

\be
H_{int}=U\sum_{i,a}\hat{n}_{ia\uparrow}\hat{n}_{ia\downarrow}+U'\sum_{i,a\ne b}
\hat{n}_{ia}\hat{n}_{ib}-J_{fd}\sum_{i,a}{\bf S}_{if}.{\bf S}_{ia} 
\ee 
\noindent describes the correlation part. In addition to $U,U'$ acting in
the $d$-manifold, we include $J_{fd}$ for the scattering
and spin polarization of the $d$-bands by well-localized $Gd\, 4f$-states 
(treated classically, i.e., neglecting spin-flip, as $S=7/2$).
As parameters relevant to \gdi, we choose $U=7.0$~eV, $U'=5.0$~eV,
and $J_{fd}=1.5$ eV, with the LSDA bandwidths between $5-6$ eV.
In addition to scattering induced by $U,\,U'$,  
$J_{fd}$ acts like a {\it classical} scattering potential
on the $d$-band states. We consider the FM phase without
additional (charge, orbital) symmetry breaking and use multi-orbital 
iterated perturbation theory (MO-IPT) \cite{LC03} to solve the multi-orbital 
impurity problem in the $d$-sector. LSDA+DMFT has been used with 
good quantitative success in many $3d$-oxides~\cite{LC03,LC03_2}. The strong 
scattering induced by $J_{fd}$ is not accounted for in multi-orbital DMFT 
and is therefore combined with the solution of the Falicov-Kimball (FK) model 
in the local approximation \cite{LC03_2}.

DMFT renormalizes the LSDA results in two steps.  First, the multi-orbital
Hartree self-energy renormalizes the relative (LSDA) band positions 
depending upon their occupations and (energy) separations.
More importantly, the frequency-dependent self-energy causes spectral-weight 
transfer across large energy scales, drastically modifying
LSDA spectra. Its importance is directly seen in photoemission
experiments on a host of correlated systems~\cite{HELD}: DMFT generally 
gives good quantitative agreement with photoemission. While LDA(LDA+U) 
generically gives correct {\it ground} states with orbital and magnetic 
order for weakly correlated(correlated) solids, its inability to describe 
{\it both} the narrowed quasiparticle bands and high-energy satellite features 
is well-documented \cite{HELD}.

\begin{figure}[tbp!]
  \begin{center}
\includegraphics[width=\columnwidth]{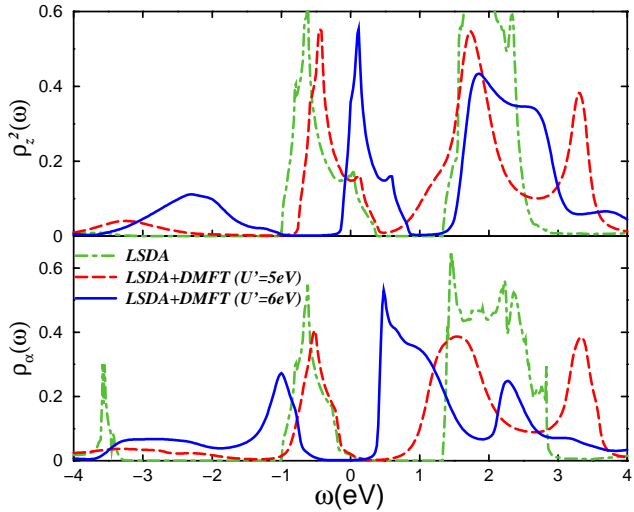}
    \caption{(Color online)
The spectral density for \dxy\, and \dx2y2\, derived bands from DMFT
calculations (lower panel). The upper panel represents the same 
for \dz2 derived bands. }
\label{mlt}
\end{center}
\end{figure}

Our DMFT results are drastically different: the Hartree self-energy 
pushes the degenerate \dx2y2, \,  \dxy\, bands down, and for $U^{\prime}=
6$ eV, moves them completely below E$_{F}$, leaving only 
the {\it narrowed} \dz2 band 
to cross E$_{F}$ (Fig.~\ref{mlt}). Smaller values of $U'$ did {\it not} 
yield this selective localization. As alluded to above, large scale spectral 
weight transfer to the Hubbard bands drastically modifies the LSDA DOS. 
In contrast to multi-orbital systems showing correlated Fermi-liquid behavior 
at low-$T$, {\it no} sharp quasiparticle peak (i.e., $Im \Sigma 
({\omega\rightarrow 0, T\rightarrow 0})\ne 0$) 
is obtained in the DOS here. This implies \cite{SI} an {\it incoherent} 
metallic behavior, with lifetimes so short that the quasiparticle 
concept loses meaning. This agrees with the high metallic resistivity 
in \gdi, which can be classified now as an orbital selective ``bad metal''. 

\begin{figure}[tbp!]
\begin{center}
\includegraphics[width=\columnwidth]{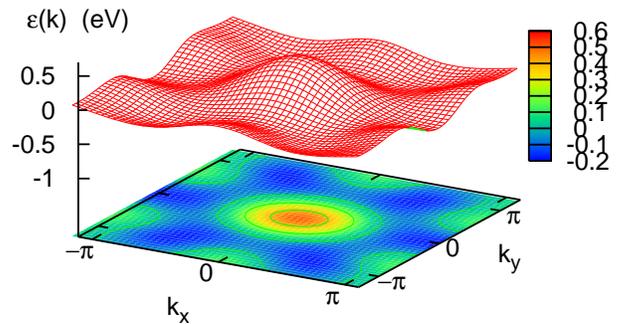}
    \caption{(Color online)
The dispersion of the LSDA+DMFT band crossing the fermi level and the
corresponding fermi surface are shown. }
\label{3dboth_rnzd}
\end{center}
\end{figure}

In fact, with Mott localization of the \dx2y2, \dxy \, bands, the problem 
effectively reduces to a FK model at low energy where the itinerant \dz2 
electrons scatter off the localized \dx2y2 and \dxy\,  
electrons. The infrared X-ray edge singularities rigorously known to
exist for the FKM in $D=\infty$~\cite{SI} give an infrared divergent
{\it local}, inter-orbital excitonic susceptibility, 
$\chi_{ab}\,(\omega)=\int dt\langle T[a_{i\sigma}^{\dag}(t)b_{i\sigma}(t);
b_{i\sigma}^{\dag}(0)a_{i\sigma}(0)\rangle e^{i\omega t}$, implying the 
existence of soft, inter-orbital ($a,(b)$= \dz2, (\dx2y2, \dxy)) 
excitonic modes. Coupling itinerant \dz2 electrons to these soft 
modes then destroys Fermi-liquid coherence leading to a ``bad metal''. Given 
finite $T=0$ entropy in this bad metal, we expect that 
a broken symmetry phase will preempt this unstable state at lower $T$.
The quasi-nested regions of the renormalised Fermi surface 
(Fig.~\ref{3dboth_rnzd}) indicate that this broken symmetry state is 
likely to be a CO state. 

We can now draw the following straightforward conclusions. Since 
the \dx2y2, \dxy \, bands are pushed below E$_{F}$ by multi-orbital DMFT, 
the small hole pockets found in the LSDA disappear in the renormalized 
Fermi surface, which now corresponds to the correlation-narrowed \dz2 band
crossing $E_F$. The DMFT orbital occupations are altered to 0.35,0.35,0.22 
for the \dx2y2, \dxy and \dz2 bands. Similar features are seen in 
DMFT work~\cite{GK} on $Na_{x}CoO_{2}$, in agreement with
ARPES~\cite{ARPES} showing no hole pockets. We predict that (i) an 
ARPES measurement performed for $80$~K $<T<T_{c}$ would show up a single 
hexagonal Fermi surface sheet, and (ii) ARPES lineshapes will be 
anomalously broad without any Fermi-liquid peaks at low energy, reflecting 
the ``bad metal'' state.

A very interesting aspect of the renormalized Fermi surface, however, 
is that, similar to dichalcogenides \cite{AEBI01}, it exhibits a built-in 
tendency to favor emergence of low-$T$ CO state of the
$\sqrt{3}\times\sqrt{3}$ type~\cite{AEBI01}. The movement of
the van-Hove-like peak in the \dz2 DOS very close to $E_{F}$ in
the LSDA+DMFT (in stark contrast to LSDA and LSDA+U results) further
corroborates this argument. Such features in Fermi surface and DOS may lead 
to unconventional CO states with gapless excitations~\cite{BAR+06}. 
Rigorous results for the existence of such CO states exist in the 
context of FK model (see below).  
 
The essential results can be understood in a simpler, effective model 
for the FM state at low energy. With electrons in \dx2y2, \dxy \, bands 
unable to hop, we introduce an {\it effective} Hamiltonian, defined by

\begin{figure}[tbp!]
\begin{center}
\includegraphics[width=\columnwidth]{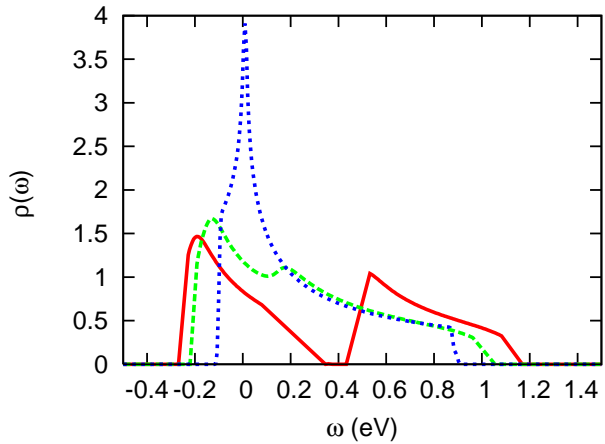}
    \caption{(Color online)
The density of states for the effective band from the model Hamiltonian 
Eqn. (3). The red and green curves are for $U_{1}=0.70$ and $0.30$eV. The
blue one is the DOS for the tight binding fit to the LSDA band that crosses 
the Fermi level. The chemical potentials are set at zero.}
\label{eff_dos}
\end{center}
\end{figure}

\bn H_{eff}&=&\sum_{k}\epsilon_{z^{2}}({\bf k})d_{z^{2},k}^{\dag}
d_{z^{2},k} +\Delta\sum_{i}(\hat{n}_{i,z^{2}}-\sum_{\alpha}\hat{n}_{i,\alpha})
\nonumber \\
&+& U_{1}\sum_{i,\alpha}\hat{n}_{i,\alpha}\hat{n}_{i,z^{2}} 
\en 

\noindent where $\alpha$=\dx2y2, \dxy. We solve $H_{eff}$ without
further approximation 
within DMFT, with the LSDA occupations ($0.31, \, 0.31,\, 0.30$ for \dxy, 
\dx2y2 and \dz2 \, orbitals). The inter-orbital correlation $U_1$ reduces 
the DOS at $E_F$ considerably (Fig.~\ref{eff_dos}) and a bad metallic 
phase, with $E_{F}$ in the low-energy pseudogap emerged. The DOS shows a 
gap for $U_1=0.70$ eV; the Fermi level, though, is never 
in the gap owing to the partial occupancy of the band.  
The {\it local} inter-orbital susceptibilities,
$\chi_{z^{2},xy}\,\, ''(\omega),\chi_{z^{2},x^{2}-y^{2}}''(\omega)$ exhibit
power-law divergences (in the $D=\infty$ FKM~\cite{SI}), yielding incoherent 
metallic behavior. 

Reduction to an effective FKM has further, concrete consequences. Since only
the \dz2 states are itinerant, the local constraint of single occupancy on
{\it each} site in $H_{eff}$ leads to the emergence of varied, 
band-filling dependent CO states. This is rigorously known \cite{UELT94} for  
FKM, and, with three (orbital) components in Hamiltonian Eqn. (3),  
we expect related CO patterns to unfold as a function of band filling.
Indications of this may already have been seen in \gdi\, \cite{private}.
We propose that such an ``ordered'' state also drives \gdi\, insulating at low
$T$. Moreover, the detailed shape of our LSDA+DMFT Fermi surface suggests 
an in-built propensity towards an unconventional CO state with an anisotropic 
gap at $E_{F}$.

The Mott-Hubbard localization of \dx2y2, \dxy\, band states found above 
leads to a local $U(1)$ invariance forcing \cite{ELITZUR75} 
$\langle d_{z^{2}}^{\dag}d_{a}\rangle = 0$ identically ($a$=\dx2y2, \dxy). 
This implies rigorous absence of inter-orbital excitonic {\it order} in 
the bad metal. Technically, this also precludes use of usual Hartree-Fock RPA 
approach for broken symmetry phases here. Interestingly enough, exactly 
this circumstance generates \cite{SI} infrared singular, 
local, excitonic modes in the uniform phase, as shown above
indicating that the FM bad metal in \gdi\, is an excitonic liquid.  
The LSDA+DMFT results do show, however, that the uniform phase 
is intrinsically unstable (given {\it soft} excitons \cite{GEO05}) to an 
inter-orbital excitonic order at low temperature. We propose that the 
unconventional CO state should also be viewed as an unconventional excitonic 
solid. We speculate that this could turn out to be a new, 
specific microscopic manifestation of a seemingly more generic 
feature in dichalcogenides \cite{AEBI01}. Suitable external tuning parameters 
(e.g., pressure, doping) may melt this state into an excitonic {\it liquid}. 
With discrete (Ising) nature of orbitals, frustration  on a triangular 
lattice may lead to partially ordered solid-like, or more exotic dimer or 
plaquette ordered phases before the unconventional state melts completely.

The structural similarity of \gdi\, to other dichalcogenides like
2H-TaS$_{2}$, 2H-TaSe$_{2}$, etc, suggests related phenomena in \gdi. In 
particular, it is possible that the unconventional CDW or 
excitonic solid (or liquid) states manifest themselves - favorable cases
would be those that are near the metal-insulator transition \cite{AEBI07}.
High pressure or chemical doping could then tune the system into a very
anomalous metallic state with strong, unconventional, excitonic {\it liquid}
correlations.  Whether such excitonic-liquid correlations also drive
unconventional superconductivity (USC) \cite{AEBI01} is an additional, 
fascinating issue in this context. 

To conclude, based on LSDA+DMFT calculations, we propose that the FM 
``metallic'' phase of \gdi\, is an excitonic liquid. At low $T$, this is argued 
to order into an unconventional excitonic CDW {\it insulator}. Our approach 
can also be used fruitfully for the investigation of other dichalcogenides 
of great 
current interest.  

LC thanks the Emmy-Noether Program of the German DFG for support. MSL
thanks the MPIPKS, Dresden for financial support.

\bibliography{tmp5}
\end{document}